\documentstyle[11pt,newpasp_crete,twoside,epsf]{article}

\begin{document}
\title{Neutron Star Masses from Arecibo Timing Observations of Five Pulsar--White Dwarf Binary Systems}

\author{David J. Nice \& Eric M. Splaver}
\affil{Physics Department, Princeton University\\ Box 708,
Princeton, NJ 08544 USA}
\author{Ingrid H. Stairs}
\affil{Department of Physics and Astronomy, University of British Columbia\\
6224 Agricultural Road, Vancouver, BC V6T 1Z1, Canada}

\begin{abstract}
We have detected relativistic and/or kinematic post-Keplerian parameters
of no fewer than five pulsar--white dwarf binary systems
in long-term timing observations at Arecibo.
We discuss the resulting constraints on pulsar and
companion masses.  The general trend is that the 
pulsar mass measurements are compatible with the canonical
value of 1.35\,M$_{\sun}$, but in most cases somewhat higher
masses are preferred.
\end{abstract}

\section{Introduction}

Evolutionary considerations suggest that pulsars in 
white dwarf--neutron star binaries might have
masses higher than the canonical value of 1.35\,M$_\odot$,
perhaps due to extended, stable mass transfer.  
Here we report on observations of five
pulsars in such binary systems.  We have measured
one or more post-Keplerian effects in each
system, leading to constraints on the pulsar masses.
(For a review of previous pulsar mass measurements,
see Thorsett \& Chakrabarty 1999.)

We observed the pulsars over several years with the
post-upgrade Arecibo telescope.  We used the Princeton Mark\,IV data
acquisition system to perform baseband sampling, coherent dedispersion,
pulse-synchronous folding, and timing (Stairs et al. 2000).  Depending
on the pulsar, we also used data collected at 
Green Bank (140 Foot telescope), Jodrell Bank, and/or Effelsberg,
as well as data from the Arecibo Berkeley Pulsar Processor and
the Princeton Mark\,III system (pre-upgrade Arecibo).  
Details will be given elsewhere.

\section{Constraints on Pulsar and Companion Mass}

\subsection{Keplerian Mass Function}

For each system, five Keplerian parameters are measured: orbital period,
$P_b$; eccentricity, $e$; angle of periastron, $\omega$; time of periastron
passage, $T_0$; and projected semi-major axis, $x=(a_1 \sin i)/c$, where $a_1$
is the semi-major axis of the pulsar orbit, $i$ is the orbital inclination
angle, and $c$ is the speed of light.  These combine to form the mass
function,
$
f_1=(m_2\sin i)^3(m_1+m_2)^{-2}
  =(2\pi/P_b)^2x^3/T_{\sun},
$
where $T_{\sun}=4.925\,\mu$s is the solar mass in time units,
and where $m_1$ and $m_2$ are the mass of the pulsar and
its companion, respectively, in solar mass units.  While $i$ is 
unknown, the constraint $\sin i\le 1$ leads to
$
m_1 < m_2^{3/2}f_1^{-1/2}-m_2.
$

\subsection{Relativistic Phenomena}

Three relativistic phenomena constrain the system masses.

{\it Precession} of an orbit can be measured if eccentricity 
is sufficiently large.  (For circular orbits, precession is completely
covariant with orbital period.)  
The rate of apsidal advance is
$
\dot{\omega} = 3(P_b/2\pi)^{-5/3}(1\!-\!e^2)^{-1}T_{\sun}^{2/3}(m_1\!+\!m_2)^{2/3},
$
so measurement of $\dot{\omega}$ yields total mass, $m_1+m_2$.  

{\it Shapiro delay}, the excess 
propagation time through the gravitational potential well of the
secondary is, for small $e$, 
$
\Delta t_s = -2\,m_2\,T_{\sun} \ln[1-\sin{i} \sin(\phi-\phi_0)],
$
where  $\phi$ is the orbital phase measured from the ascending node, $\phi_0$.
For edge-on orbits (large $i$), detection of Shapiro delay leads to measurement
of $m_2$ and $\sin i$.  For face-on orbits (small $i$), Shapiro delay
is covariant with projected semi-major axis and cannot be detected;
this non-detection can be used to exclude
high-inclination, low-companion-mass orbits.

{\it Orbital decay} due to 
gravitational radiation causes $P_b$ to change at a rate
$
\dot P_b  =  -(192\pi/5)
  (P_b/2\pi)^{-5/3}
  (1\!+\!\frac{73}{24} e^2\!+\!\frac{37}{96} e^4)
  (1\!-\!e^2)^{-7/2}\,T_\odot^{5/3}\, m_1\, m_2\, (m_1\!+\!m_2)^{-1/3}.
$
This can be measured for sufficiently small orbits (low $P_b$).

\subsection{Kinematic $\dot{x}$}

Motion of a binary system relative to the Sun induces a change in 
the inclination angle of the orbit, and hence a change in the 
projected semi-major axis,
$\dot{x}/x=-\mu\,\cot i\,\sin\theta$,
where $\mu$ is the (measured) proper motion and $\theta$ is the
({\it a priori} unknown) difference between the position angle
of proper motion and the position angle of the ascending node
(Kopeikin 1996; see also Nice, Splaver, \& Stairs 2001).
Measurement of $\dot{x}$ alone does not yield a value of $i$,
but the constraint $|\sin\theta|\le 1$ leads to
$i<\tan^{-1}(\mu x/\dot{x})$,
giving a firm upper limit on $i$.

\subsection{Orbital Period--Core Mass relation.}

Studies of low- and intermediate-mass binary evolution predict
a nearly unique relation between $P_b$ and $m_2$
in pulsar--helium white dwarf binaries (Podsiadlowski,
Rappaport, \& Pfahl 2002; Tauris \& Savonije 1999; Rappaport et al.
1995).  The $P_b\!-\!m_2$ relation may be used to constrain the system
masses;  alternatively, measured system masses may be
used to test the evolution theory.

\section{Timing Analysis and Results}

Table 1 summarizes the post-Keplerian 
phenomena detected in each of five pulsar binaries.  For each pulsar, we used 
{\sc tempo}\footnote{http://pulsar.princeton.edu/tempo}
to calculate the goodness-of-fit of 
timing solutions over a uniform grid of $\cos i$ and $m_2$ values. In each
timing run, relativistic parameters appropriate to the given
$\cos i$ and $m_2$ were calculated and held constant, while all
other parameters were allowed to vary.  The gray regions in figures 1 and 2
show combinations of $\cos i$ and $m_2$ that gave acceptable
fits at the 68\% and 95\% confidence levels (inner and outer 
contours).\footnote{For J0621+1002, the contours are at 39\% and 86\% confidence levels.}

For each pulsar (except J2019+2425), the grids of timing
runs were used to calculate the probability distribution function (pdf) for the
underlying value of $m_1$.  Briefly, the $\Delta\chi^2$ value from each grid
point was translated into a probability and, after suitable normalization, the
probabilities associated with a given range of $m_1$ were summed, resulting
in a pdf for $m_1$.  The limits on $m_1$ from each pdf are given in
table 2.  See Splaver et al. 2002 for more details on this procedure.

\begin{table}[t]
\caption{Pulsar Parameters and Post-Keplerian Detections}

\begin{tabular}{lccccccc}
\tableline
\tableline
\multicolumn{1}{c}{Pulsar} & $P_b$    &   $e$        & $f_1$    & 
                     \multicolumn{4}{c}{Post-Keplerian Parameters} \\
\cline{5-8}
       &  (Days)  &              &    (M$_{\sun}$) &  
                     $\dot{\omega}$ & Shapiro & $\dot{P_b}$\rule{0pt}{12pt} & $\dot{x}$ \\
\tableline
J0621+1002 & \phantom{0}8.32 & 0.002\,457 & 0.0270 & $\surd$ & limit   &         &      \\
J0751+1807 & \phantom{0}0.26 & 0.000\,003 & 0.0010 &         & maybe?  & soon!   &         \\
J1713+0747 &           67.83 & 0.000\,075 & 0.0079 &         & $\surd$ &         & $\surd$     \\
B1855+09   &           12.33 & 0.000\,022 & 0.0056 &         & $\surd$ &         &      \\
J2019+2425 &           76.51 & 0.000\,111 & 0.0107 &         & limit   &         & $\surd$     \\
\tableline
\tableline
\end{tabular}
\end{table}

\subsection{PSR\,J0621+1002}

The relatively high eccentricity of this system (still only $e=0.0025$!)
is sufficient to break the covariance between $P_b$ and $\dot{\omega}$ in
the timing fit.  We measure $\dot{\omega}=0.0116\pm 0.0008^\circ\,{\rm yr}^{-1}$ 
($1\sigma$), yielding $m_1+m_2=2.8\pm0.3\,{\rm M}_{\odot}$.  This corresponds
to the strip between dashed lines in figure 1a.  We do not detect Shapiro
delay, so orbits with high $i$ are excluded, closing off
the gray region towards the lower right portion of figure 1a.
The low eccentricity
of the system, along with the lack of a detectable optical companion, provides
strong evidence the companion is a white dwarf, so $m_2<1.4\,{\rm M}_{\odot}$.
As figures 1a and 1b show, the orbit is relatively close to face on,
and the pulsar and companion masses are relatively high.
Further discussion can be found in Splaver et al. 2002.

\subsection{PSR\,J0751+1807}

This binary has a very short orbital period, only 6.3 hours.  
The relativistic decay rate is expected to be 
around $\dot{P_b}\sim-4\times 10^{-14}$,
depending on the system masses (see figure 2a for
representative calculations).   The current data give 
$\dot{P_b}=(2\pm 8)\times 10^{-14}$, tantalizingly
close to the expected value.  Results of a full relativistic 
timing analysis (figure 2a) show that higher values of $i$ are
preferred, implying a marginal detection of Shapiro delay.
The pulsar mass is not yet   
measured, though there is a strong upper limit $m_2<2.07$\,M$_{\sun}$.

We anticipate that future observations will substantially
reduce the uncertainty of $\dot{P_b}$.  (For a pulsar observed
uniformly over time span $t$, the uncertainty of $\dot{P_b}$ 
scales as $t^{-2.5}$.) This will be the first 
detection of relativistic decay outside of double-neutron-star
binaries.

\begin{table}[t]
\caption{Mass Measurements from Timing Analysis}

\begin{tabular}{lccc}
\tableline
\tableline
\multicolumn{1}{c}{Pulsar} & \multicolumn{3}{c}{Pulsar Mass, $m_1$ (M$_{\sun}$)} \\
\cline{2-4}
       & Best Value & 68\% confidence & 95\% confidence \\
\tableline
J0621+1002 & 1.70 & 1.41$-$2.02 & 1.07$-$2.29 \\
J0751+1807 & 0.55 & 0.19$-$1.25 & 0.02$-$2.07 \\
J1713+0747 & 1.85 & 1.49$-$2.26 & 1.23$-$2.83 \\
B1855+09   & 1.57 & 1.46$-$1.69 & 1.37$-$1.82 \\
\tableline
\tableline
\end{tabular}
\end{table}

\subsection{PSR\,J1713+0747}

This pulsar in a wide binary has a strong, sharp pulse, allowing us
to achieve timing precision of 200$-$400\,ns in a single
observation.  Camilo, Foster, \& Wolszczan (1994)
first reported the detection of Shapiro delay.  Our data
give substantially improved constraints on the system masses
and inclination (figure 2b).  The timing data still allow a
wide range of pulsar masses (table 2), but 
the overlap between parameters allowed by the timing 
data and the range of $m_2$
predicted by the $P_b\!-\!m_2$ relation of Tauris \& Savonije (1999)
gives a tighter constraint on $m_1$, $1.46\!-\!1.61\,{\rm M}_{\odot}$ (figure 2b).
(The binary evolution tracks of
Podsiadlowski et al. 2002 give very similar predictions.)

\subsection{PSR\,B1855+09}

Our new data only modestly augments that of
Kaspi, Taylor, \& Ryba (1994) and earlier work.  Shapiro delay
gives tight constraints on $m_2$ and $i$ (figure 2c), 
which are in excellent agreement with the $P_b\!-\!m_2$
relation.

\subsection{PSR\,J2019+2425}

No relativistic effects are detected.
The observed ${\dot{x}/x}=1.3\pm0.2\times 10^{-15}$\,s$^{-1}$
constrains $i<72^\circ$.  Shapiro delay
is not detected, also constraining the system away
from high $i$ (figure 2d). Little can be said
about the system masses from timing alone.  However, 
if $m_2$ is within the range allowed
the $P_b\!-\!m_2$ relation (horizontal strip in figure 2d),
the neutron star mass must be $m_1<1.51$\,M$_\odot$.
Further analysis of this system is given in 
Nice, Splaver, \& Stairs 2001.

\section{Conclusion}

Table 2 shows substantial constraints on neutron star mass $m_1$ from timing 
alone for three of the five pulsars, J0621+1002, J1713+0747, and B1855+09.
In each case, the 68\% confidence limits on $m_1$ are above the canonical
neutron star mass of 1.35\,M$_\odot$, although the 95\% ranges
all encompass this value.  PSR J0437$-$4715 is similar, with
$m_1=1.58\pm0.18\,$M$_{\odot}$ (van Straten et al. 2001).
The evidence is mounting that some
pulsars are heavier than 1.35\,M$_\odot$, though this cannot yet
be considered conclusive.  The $P_b\!-\!m_2$ relation is in good
agreement with the timing data and with  the
relatively high pulsar masses.

The most exciting prospect for the future is PSR~J0751+1807.  Although
its mass is only weakly constrained, especially on the lower end, 
great advances will be made as the decay of this pulsar's orbit is
measured in coming years.

\begin{figure}

\plottwo{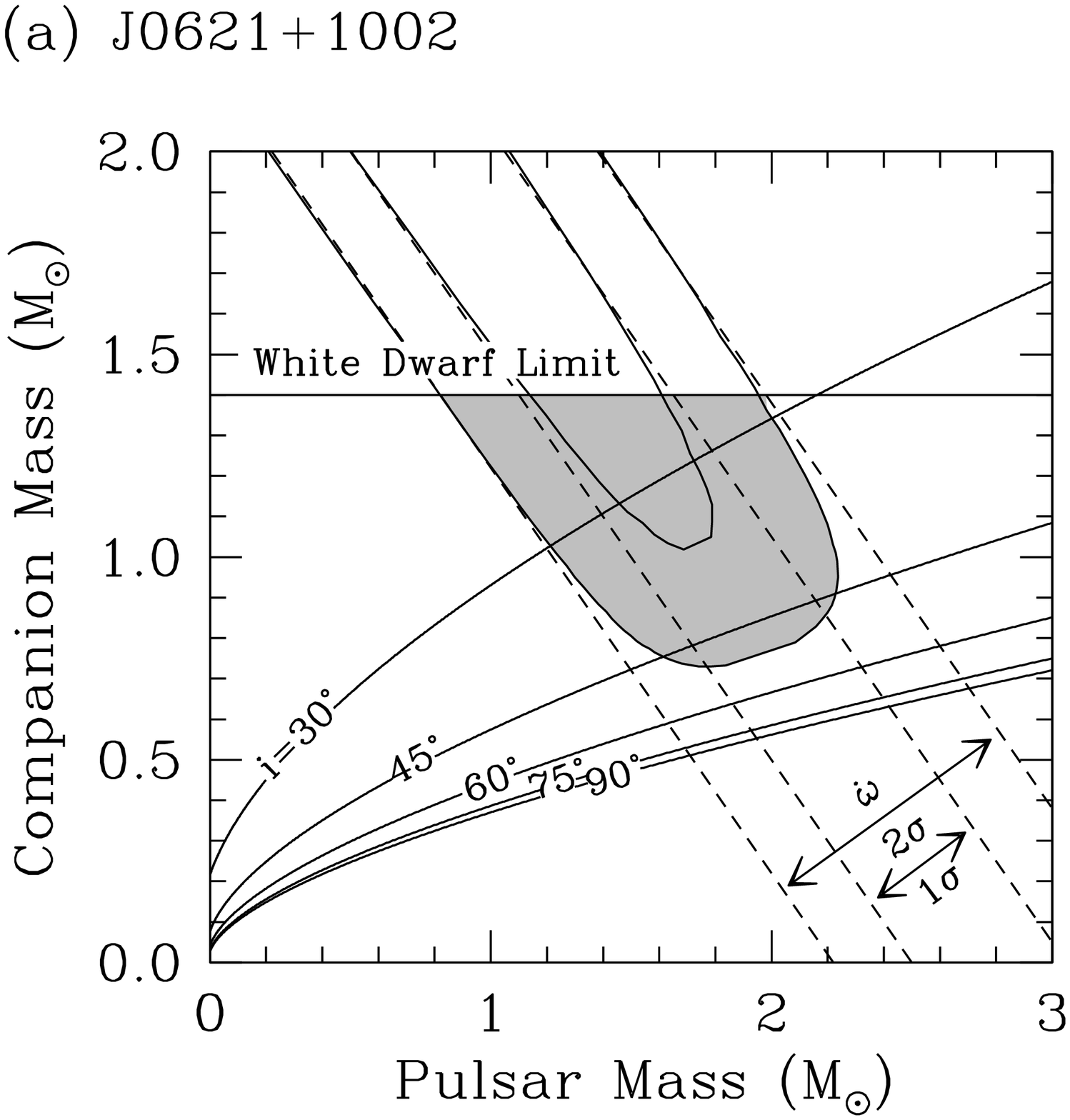}{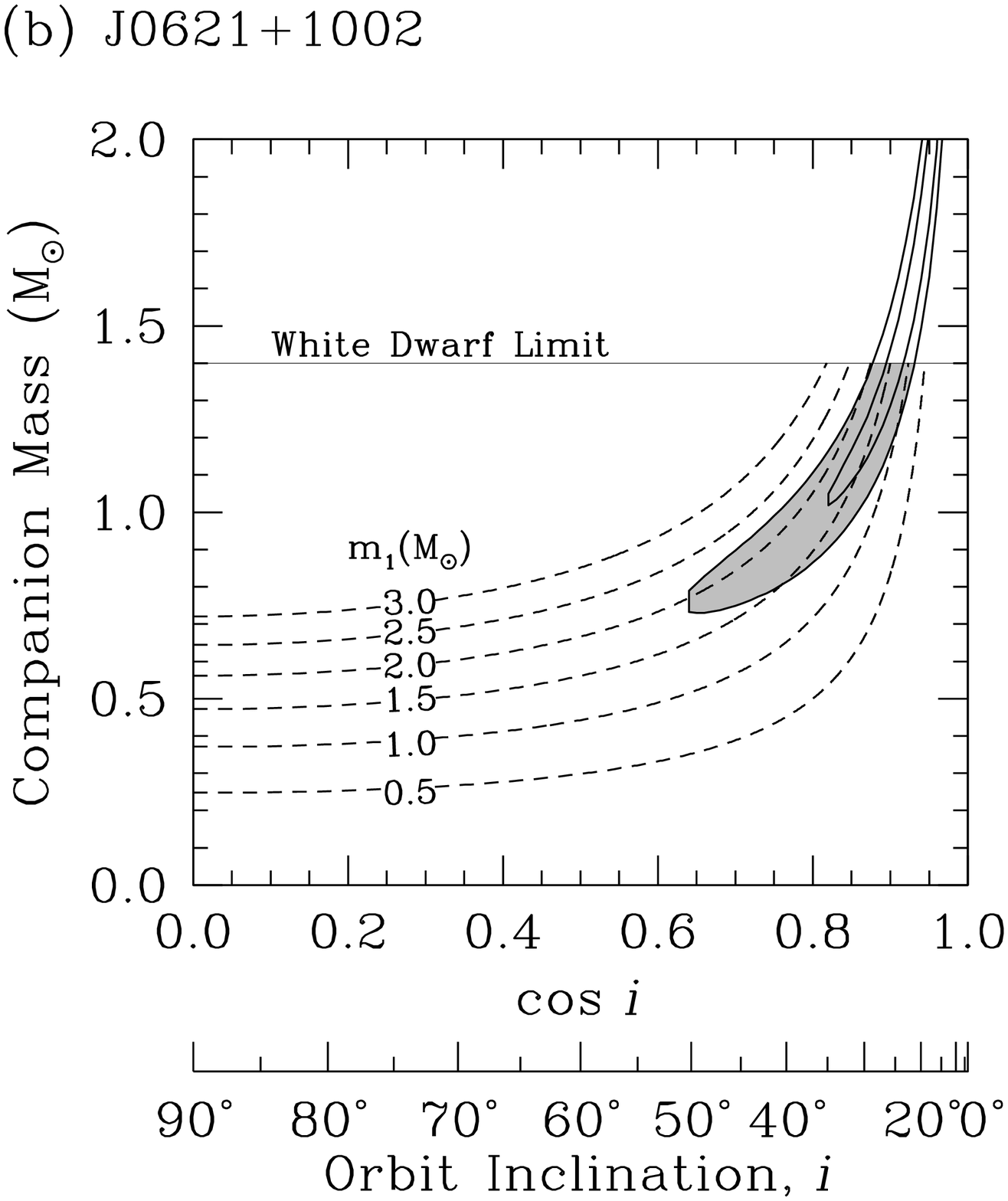}

\caption{Constraints on PSR J0621+1002 parameters.  See text.}
\end{figure}

\begin{figure}

\plottwo{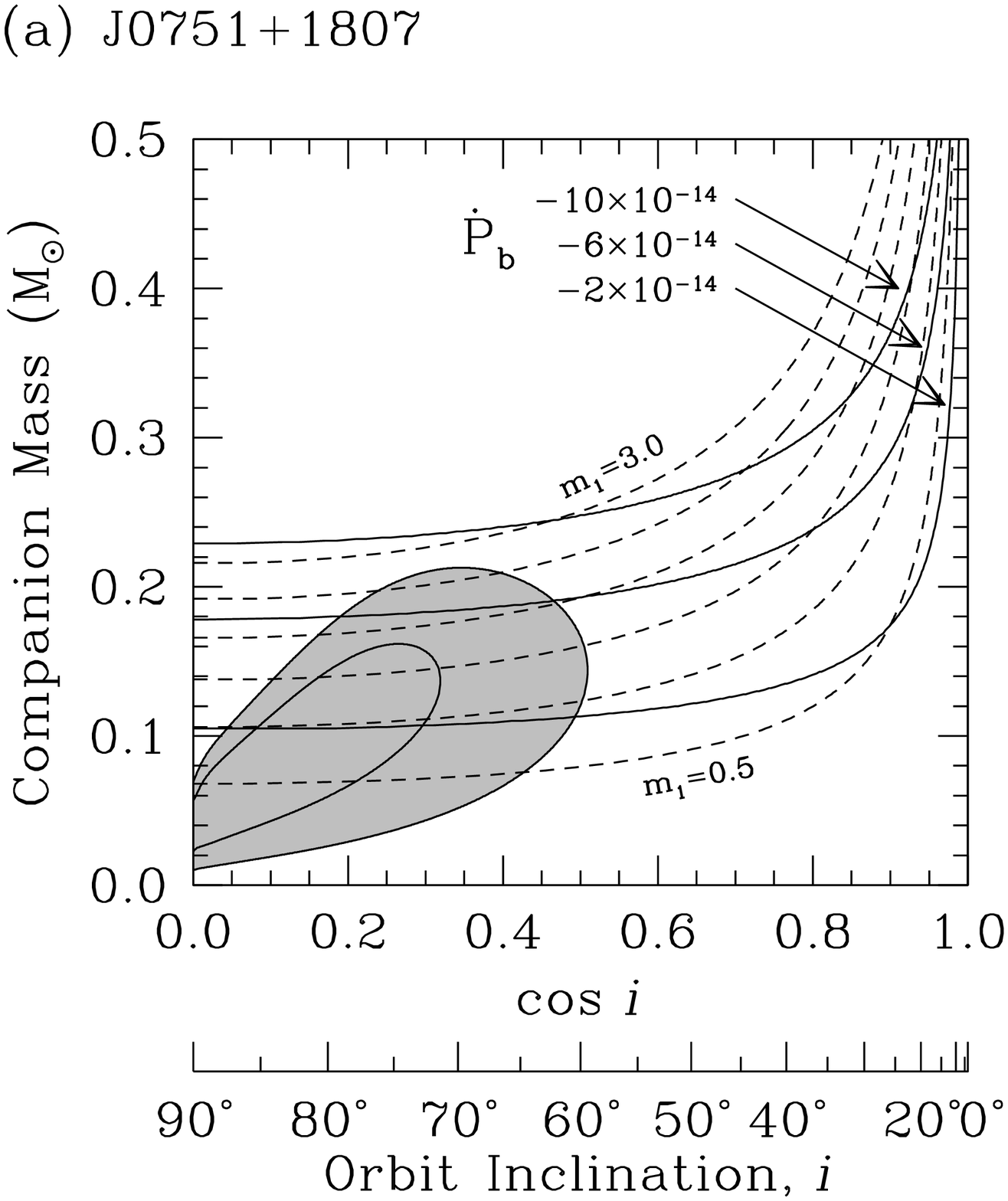}{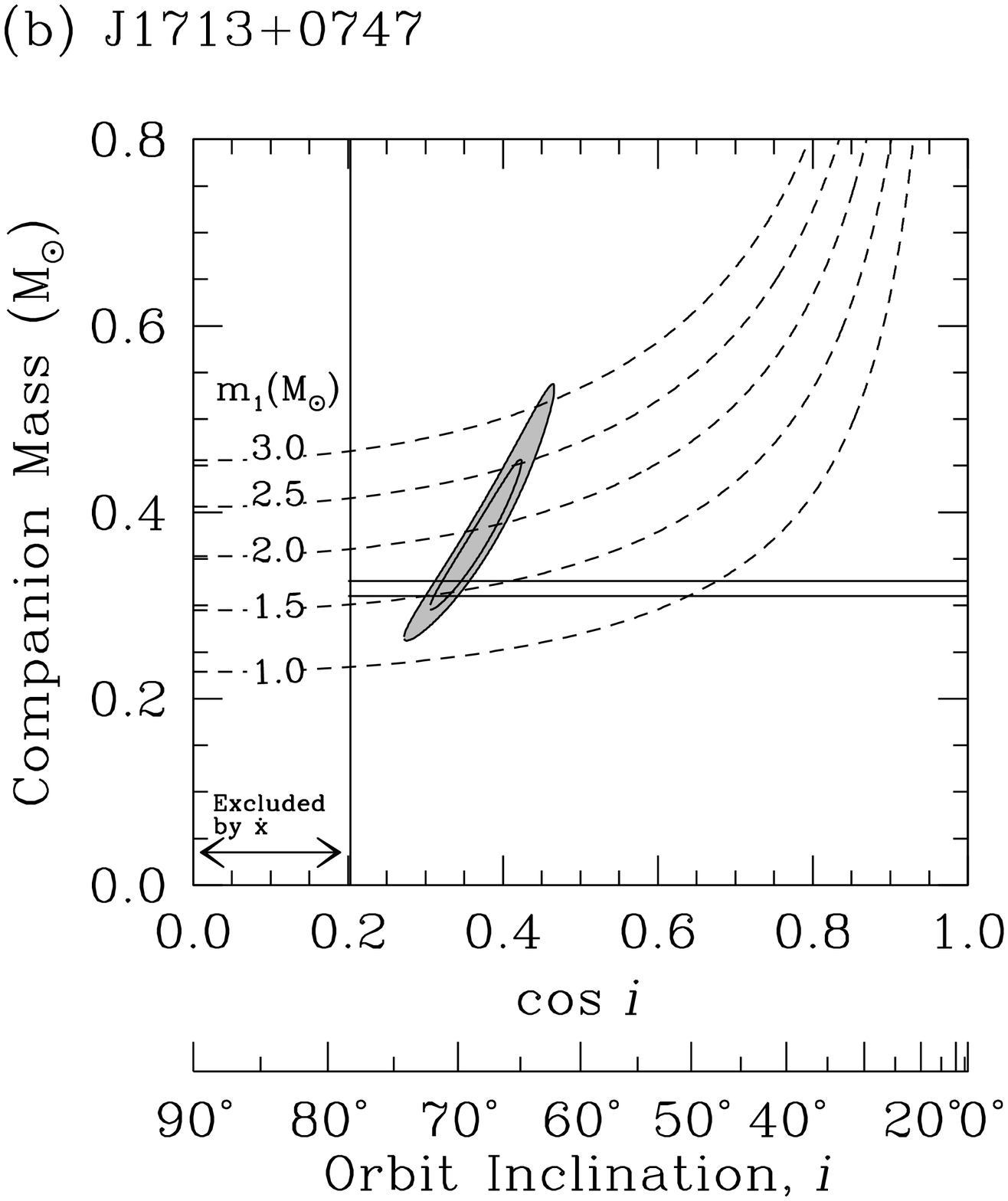}

\plottwo{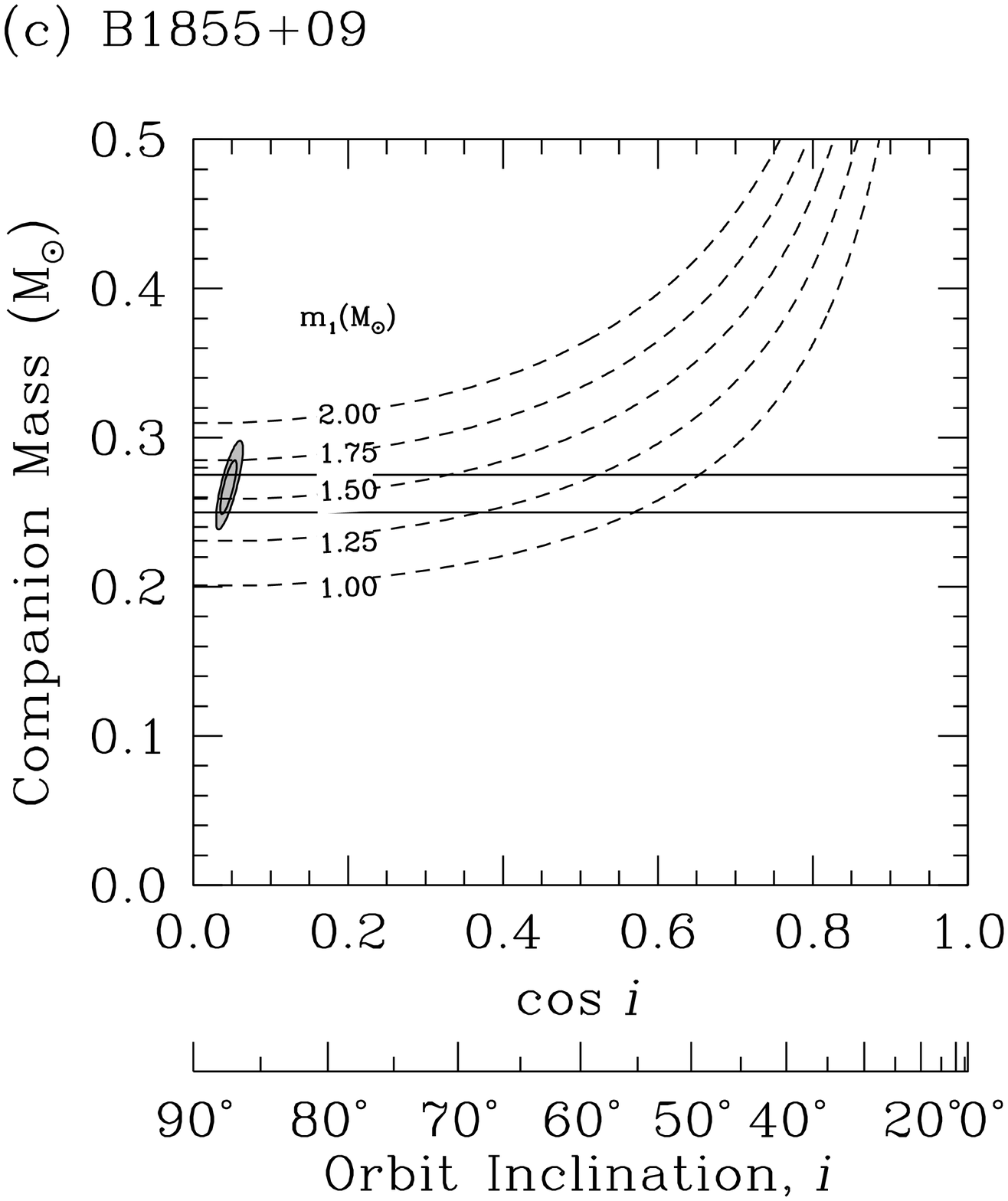}{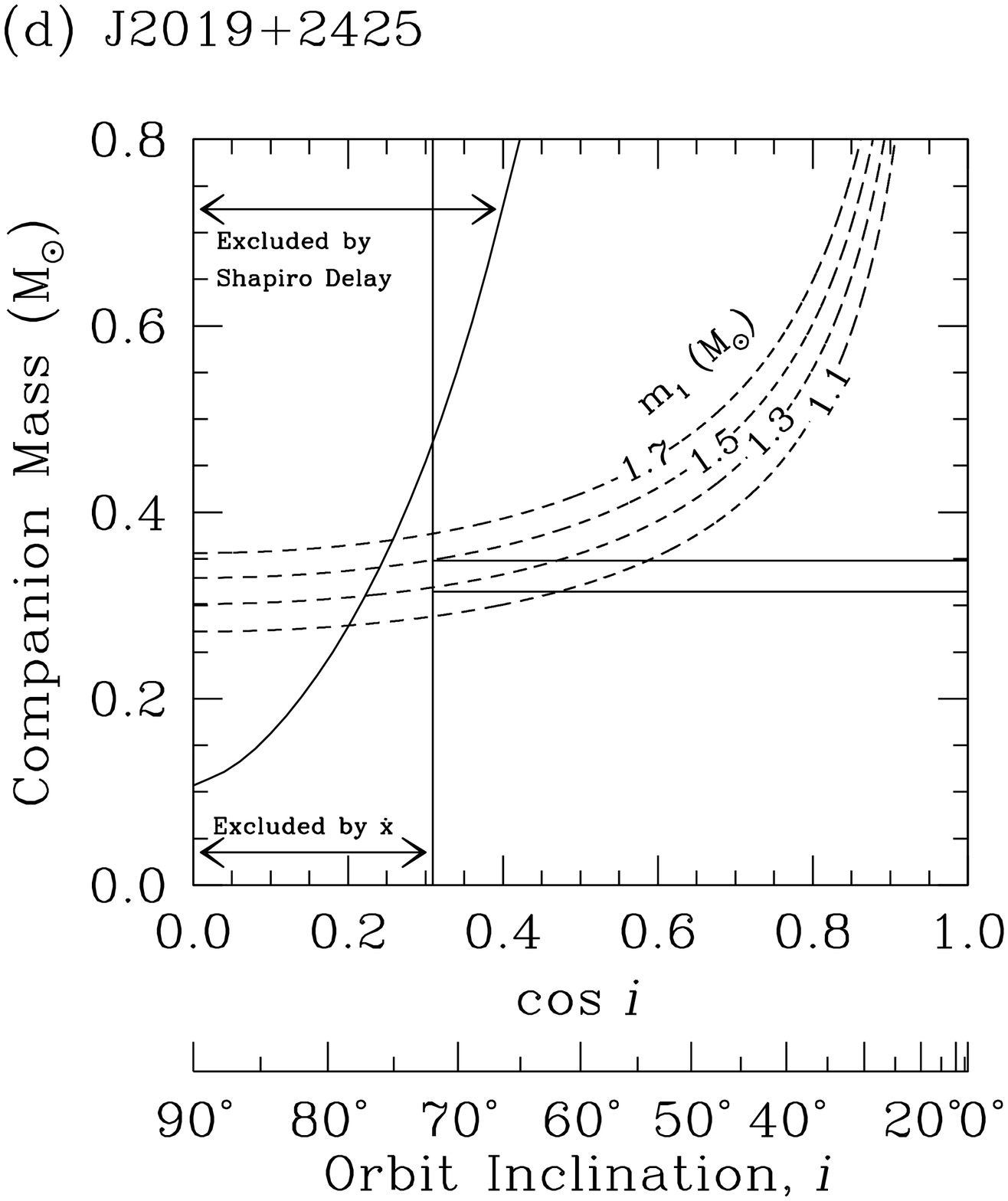}

\caption{Constraints on $\cos i$ and $m_2$.
Dashed lines show representative values of $m_1$.
Gray regions denote the area allowed by the data.  When present,
a vertical line indicates the largest $i$ allowed by a measured
$\dot{x}$, and horizontal lines indicate
the range of $m_2$ predicted by the $P_b\!-\!m_2$ relation.
(a) PSR~J0751+1807.  Solid lines show representative values of $\dot{P_b}$.  
(b) PSR~J1713+0747.
(c) PSR~B1855+09.  
(d) PSR~J2019+2425.  The curved solid line delimits the region in the upper left excluded
by the non-detection of Shapiro delay.
}
\end{figure}

\acknowledgements

Portions of this work are collaborations with Zaven
Arzoumanian, Don Backer, Fernando Camilo, Michael Kramer, Andrea Lommen,
Oliver L\"{o}hmer, and Andrew Lyne.  We are indebted to Joe
Taylor and Steve Thorsett for development work on the Princeton Mark IV
system and to Kiriaki Xilouris and Dunc Lorimer for efforts 
in advancing pulsar astronomy on the post-upgrade Arecibo telescope.
The Arecibo Observatory is operated by Cornell University under a
cooperative agreement with the National Science Foundation.  Pulsar
research at Princeton University is supported by the NSF.
IHS is supported by an NSERC UFA and Discovery Grant.

\end{document}